\documentstyle[12pt,epsf,rotate]{article}

\def\epe{\varepsilon'/\varepsilon}
\newcommand{\mt}{m_{\rm t}}
\newcommand{\mtb}{\overline{m}_{\rm t}}
\newcommand{\mc}{m_{\rm c}}
\newcommand{\ms}{m_{\rm s}}

\newcommand{\mb}{m_{\rm b}}
\newcommand{\mw}{M_{\rm W}}
\newcommand{\mz}{M_{\rm Z}}
\newcommand{\gev}{\, {\rm GeV}}
\newcommand{\mev}{\, {\rm MeV}}

\newcommand{\Lms}{\Lambda_{\overline{\rm MS}}}

\newcommand{\RE}{{\rm Re}}
\newcommand{\IM}{{\rm Im}}

\newcommand{\vcb}{|V_{cb}|}
\newcommand{\vtd}{|V_{td}|}
\newcommand{\vub}{|V_{ub}/V_{cb}|}

\newcommand{\bea}{\begin{eqnarray}}
\newcommand{\eea}{\end{eqnarray}}
\newcommand{\bd}{\begin{displaymath}}
\newcommand{\ed}{\end{displaymath}}

\begin{document}
\begin{flushright}
 TUM-HEP-255/96 \\
 MPI-PhT/96-87 \\
hep-ph/9609324 \\
 September 1996
\end{flushright}
\vskip1truecm
\centerline{\Large\bf  Theoretical Review of K-Physics
   \footnote[2]{\noindent Theoretical Summary Talk given at
 the "Workshop on K-Physics", Orsay, May 30 - June 4, 1996, to
appear in the proceedings.\\
   Supported by the German
   Bundesministerium f\"ur Bildung and Forschung under contract
   06 TM 743 and DFG Project Li 519/2-1.}}
\vskip1truecm
\centerline{\bf Andrzej J. Buras}
\bigskip
\centerline{\sl Technische Universit\"at M\"unchen, Physik Department}
\centerline{\sl D-85748 Garching, Germany}
\vskip0.6truecm
\centerline{\sl Max-Planck-Institut f\"ur Physik}
\centerline{\sl  -- Werner-Heisenberg-Institut --}
\centerline{\sl F\"ohringer Ring 6, D-80805 M\"unchen, Germany}
\vskip1truecm
\thispagestyle{empty}
\centerline{\bf Abstract}

We review several aspects of K-Physics:
i) Main targets of the field,
ii) The theoretical framework for K-decays, 
iii) Standard analysis of the unitarity triangle, 
iv) $\varepsilon'/\varepsilon$,
v) Rare and CP violating K-decays, vi) Comparision of the potentials of
$ K \to \pi\nu\bar\nu $ and CP-B asymmetries.
vii) Some aspects of the physics beyond the Standard Model.

\newpage
\setcounter{page}{1}

\section{Introduction}

The theoretical review of K-physics presented below is based on 
the theoretical summary talk which
I have given at the K-physics workshop held in Orsay this summer. The 
experimental summary has been presented by Bruce Winstein.

This workshop has shown very clearly a great potential of K-physics
in testing the Standard Model, testing its possible extensions and 
searching for exotic phenomena such as lepton number violation, CPT and
Quantum Mechanics violations. In view of space limitations not all
topics can be presented here and I will frequently refer to other
talks contained in these proceedings.

I have organized the material as follows:

Section 2 gives a " Grand View" of the field, discussing 
its most important targets, recalling the CKM matrix and the unitarity
triangle and presenting briefly the theoretical framework.

Section 3 discusses the by now standard analysis of the unitarity triangle
(UT). 

Section 4 summarizes the present status of  $\epe$.

Section 5 summarizes the present status of the four stars in the field of 
rare K-decays:  $K_L\to\pi^o e^+ e^-$, 
 $K^+\to\pi^+\nu\bar\nu$, $K_L\to\pi^o\nu\bar\nu$ and $K_L\to \mu\bar\mu$.
This section ends with a classification of  K- and B-decays from the point
of view of theoretical cleanliness.

Section 6 compares the potentials of CP asymmetries in B-decays and of
the very clean decays $K^+\to\pi^+\nu\bar\nu$ and $K_L\to\pi^o\nu\bar\nu$ 
in determining the parameters of the CKM matrix. Here
necessarily a short discussion of CP violation in B-decays 
will be given.

Section 7 offers a brief look beyond the Standard Model.

Section 8  gives a very short outlook. 

Section 9 contains some remarks on this workshop.

\section{Grand View}
\subsection{Main Targets of K-Physics}
Let us list the main targets of K-Physics:
\begin{itemize}
\item
The parameters of the CKM matrix. In particular: the parameters $\lambda$,
$\eta$, $\varrho$, the element $|V_{td}|$ and $\sin 2\beta$,
\item
CP violation and rare decays in the Standard Model,
\item
Low energy tests of QCD, tests and applications of non-perturbative
methods such as: lattice, chiral perturbation theory, 1/N expansion,
QCD sum rules, hadronic sum rules,
\item
Physics beyond the Standard Model such as supersymmetry, left-right
symmetry, charged higgs scalars, leptoquarks, lepton number violations
etc.,
\item
Truly exotic physics related to CPT violations and Quantum Mechanics
violations.
\end{itemize}
All of these targets have been discussed at this workshop. In particular
searches for CPT violations and tests of Quantum Mechanics for which
DA${\Phi}$NE is clearly an excellent machine, have been elaborated
on by Cline, Di Domenico, Ellis, Tsai, Kostelecky and Huet. CPT tests
outside the K system have been discussed by Gabrielse and Okun. Finally
very interesting results on CP, T and CPT tests at CPLEAR and phase
measurements at CERN and Fermilab have been presented by Le Gac and
Pavlopoulos and vigorously discussed at the round table discussion on
the phase measurements. Related issues have been presented by Khalfin. 
Since these aspects are already summarized by
Bruce Winstein I will not include them in my talk. Let me then move to the
CKM matrix which is central for this field.

\subsection{The CKM Matrix and the Unitarity Triangle}
An important target of particle physics is the determination
 of the unitary $3\times 3$ Cabibbo-Kobayashi-Maskawa
matrix \cite{CAB,KM} which parametrizes the charged current interactions of
 quarks:
\begin{equation}\label{1j}
J^{cc}_{\mu}=(\bar u,\bar c,\bar t)_L\gamma_{\mu}
\left(\begin{array}{ccc}
V_{ud}&V_{us}&V_{ub}\\
V_{cd}&V_{cs}&V_{cb}\\
V_{td}&V_{ts}&V_{tb}
\end{array}\right)
\left(\begin{array}{c}
d \\ s \\ b
\end{array}\right)_L
\end{equation}
The CP violation in the standard model is supposed to arise
from a single phase in this matrix.
It is customary these days to express the CKM-matrix in
terms of four Wolfenstein parameters 
\cite{WO} $(\lambda,A,\varrho,\eta)$
with $\lambda=\mid V_{us}\mid=0.22 $ playing the role of an expansion 
parameter and $\eta$
representing the CP violating phase:
\begin{equation}\label{2.75} 
V_{CKM}=
\left(\begin{array}{ccc}
1-{\lambda^2\over 2}&\lambda&A\lambda^3(\varrho-i\eta)\\ -\lambda&
1-{\lambda^2\over 2}&A\lambda^2\\ A\lambda^3(1-\varrho-i\eta)&-A\lambda^2&
1\end{array}\right)
+O(\lambda^4)
\end{equation}
Because of the
smallness of $\lambda$ and the fact that for each element 
the expansion parameter is actually
$\lambda^2$, it is sufficient to keep only the first few terms
in this expansion. 

Following \cite{BLO} one can define the parameters
$(\lambda, A, \varrho, \eta)$ through
\begin{equation}\label{wop}
s_{12}\equiv\lambda \qquad s_{23}\equiv A \lambda^2 \qquad
s_{13} e^{-i\delta}\equiv A \lambda^3 (\varrho-i \eta)
\end{equation}
where $s_{ij}$ and $\delta$ enter the standard exact 
parametrization \cite{PDG}  of the CKM
matrix. This specifies the higher orders terms in (\ref{2.75}).

The definition of $(\lambda,A,\varrho,\eta)$ given in (\ref{wop})
is useful because it allows to improve the accuracy of the
original Wolfenstein parametrization in an elegant manner. In
particular
\begin{equation}\label{CKM1}
V_{us}=\lambda \qquad V_{cb}=A\lambda^2
\end{equation}
\begin{equation}\label{CKM2}
V_{ub}=A\lambda^3(\varrho-i\eta)
\qquad
V_{td}=A\lambda^3(1-\bar\varrho-i\bar\eta)
\end{equation}
where
\begin{equation}\label{3}
\bar\varrho=\varrho (1-\frac{\lambda^2}{2})
\qquad
\bar\eta=\eta (1-\frac{\lambda^2}{2})
\end{equation}
turn out \cite{BLO} to be excellent approximations to the
exact expressions.

\begin{figure}[hbt]
\vspace{0.10in}
\centerline{
\epsfysize=2.0in
\epsffile{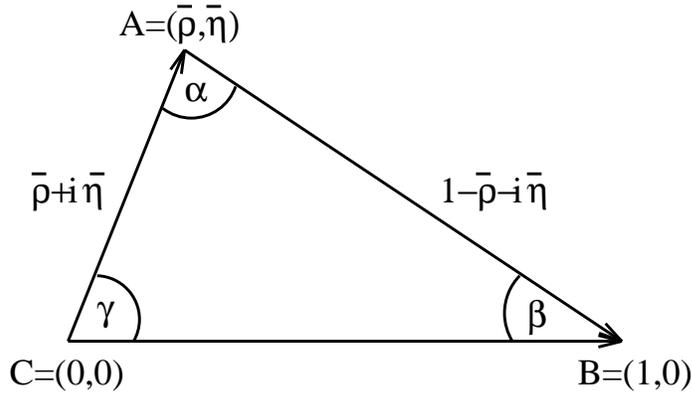}
}
\vspace{0.08in}
\caption[]{
Unitarity Triangle.
\label{fig:utriangle}}
\end{figure}

A useful geometrical representation of the CKM matrix is the unitarity 
triangle obtained by using the unitarity relation
\begin{equation}\label{2.87h}
V_{ud}V_{ub}^* + V_{cd}V_{cb}^* + V_{td}V_{tb}^* =0,
\end{equation}
rescaling it by $\mid V_{cd}V_{cb}^\ast\mid=A \lambda^3$ and depicting
the result in the complex $(\bar\rho,\bar\eta)$ plane as shown
in fig. 1. The lenghts CB, CA and BA are equal respectively to 1,
\begin{equation}\label{2.94a}
R_b \equiv  \sqrt{\bar\varrho^2 +\bar\eta^2}
= (1-\frac{\lambda^2}{2})\frac{1}{\lambda}
\left| \frac{V_{ub}}{V_{cb}} \right|
\qquad
{\rm and}
\qquad
R_t \equiv \sqrt{(1-\bar\varrho)^2 +\bar\eta^2}
=\frac{1}{\lambda} \left| \frac{V_{td}}{V_{cb}} \right|.
\end{equation}

The triangle in fig. 1, $\mid V_{us}\mid$ and $\mid V_{cb}\mid$
give the full description of the CKM matrix. 
Looking at the expressions for $R_b$ and $R_t$ we observe that within
the standard model the measurements of four CP
{\it conserving } decays sensitive to $|V_{us}|$, $\vcb$,   
$| V_{ub}| $ and $ |V_{td}|$ can tell us whether CP violation
($\eta \not= 0$) is predicted in the standard model. 
This is a very remarkable property of
the Kobayashi-Maskawa picture of CP violation: quark mixing and CP violation
are closely related to each other. 

There is of course the very important question whether the KM picture
of CP violation is correct and more generally whether the standard
model offers a correct description of weak decays of hadrons. In order
to answer these important questions it is essential to calculate as
many branching ratios as possible, measure them experimentally and
check if they all can be described by the same set of the parameters
$(\lambda,A,\varrho,\eta)$. In the language of the unitarity triangle
this means that the various curves in the $(\bar\varrho,\bar\eta)$ plane
extracted from different decays should cross each other at a single point
as shown in fig. 2.
Moreover the angles $(\alpha,\beta,\gamma)$ in the
resulting triangle should agree with those extracted one day from
CP-asymmetries in B-decays. More about this below.

\begin{figure}[hbt]
\vspace{0.10in}
\centerline{
\epsfysize=6.5in
\rotate[r]{
\epsffile{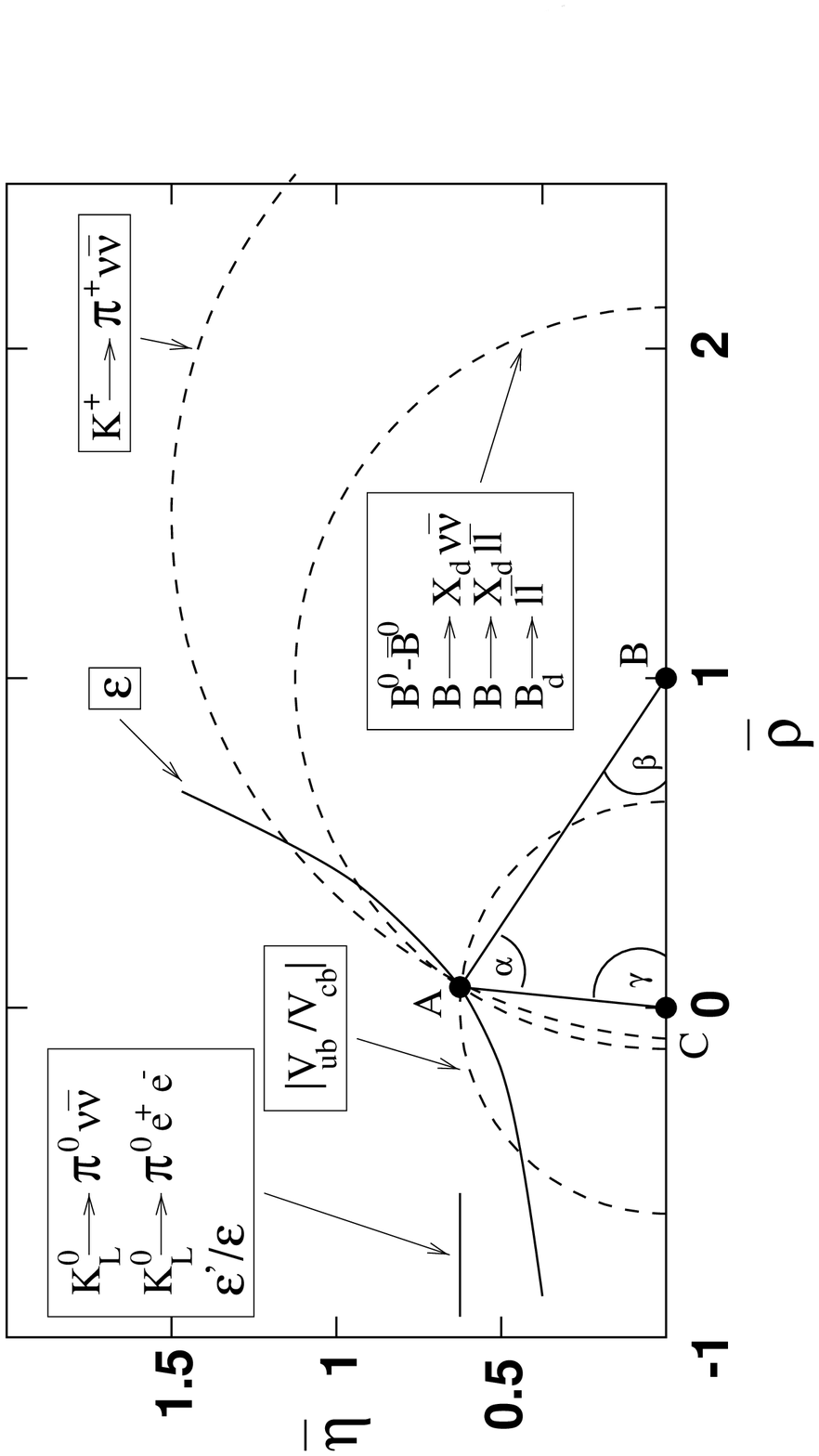}
}}
\vspace{0.08in}
\caption[]{
The ideal Unitarity Triangle. For artistic reasons the value of
$\bar\eta$ has been chosen to be higher than the fitted central value
$\bar\eta\approx 0.35.$
\label{fig:2011}}
\end{figure}

Since the CKM matrix is only a parametrization of quark mixing and 
of CP violation and does not offer the explanation of these two
very important phenomena, many physicists hope that a new physics
while providing a dynamical origin of quark mixing and CP violation will
also change the picture given in fig. 2. That is, the different curves
based on standard model expressions, will not cross at a single point
and the angles $(\alpha,\beta,\gamma)$ 
extracted one day from
CP-asymmetries in B-decays will disagree with the ones determined from
rare K and B decays.

Clearly the plot in fig. 2 is highly idealized because in order
to extract such nice curves from various decays one needs perfect
experiments and perfect theory. We will see in section 6 that 
for certain decays such a picture is not fully unrealistic.
Generally however the task of extracting the unitarity triangle
from the experimental data is not easy.
 In order to understand this we have to discuss the present 
theoretical framework.

\subsection{Theoretical Framework}
The basic problem in the calculation of branching ratios for K decays
and other physical observables in the K system is related to strong 
interactions. Although due to  
the smallness
of the effective QCD coupling at short distances, the gluonic
contributions at scales ${\cal O} (\mw, \mz, \mt)$ can be calculated
within the perturbative framework, the fact that
K mesons are $ q\bar q$ bound states forces us to consider  QCD at 
long distances as well.
 Here we have to rely on existing non-perturbative
methods, discussed briefly below, which are not yet very powerful
at present.

The question then is whether we could somehow divide the problem
into two parts: the short distance part, under control already today,
and the long distance part which hopefully will be fully under control
when our non-perturbative tools improve. One could even hope that
in certain decays the non-perturbative contributions could be
measured and subsequently used in other decays so that one could
make predictions in some cases already today without the necessity
for non-perturbative calculations.

The Operator Product Expansion (OPE) combined with the renormalization group
approach can be regarded as a mathematical formulation of the strategy
outlined above. This framework 
brings in local operators $Q_i$ which govern ``effectively''
the transitions in question and 
the amplitude for an {\it exclusive} decay $M\to F$ 
is written as
\begin{equation}\label{OPE}
 A(M \to F) = \frac{G_F}{\sqrt 2} {\rm V_{CKM}} \sum_i
   C_i (\mu) \langle F \mid Q_i (\mu) \mid M \rangle 
\end{equation}
where $M$ stands for the decaying meson, $F$ for a given final state and
${\rm V_{CKM}}$ denotes the relevant CKM factor.
$ Q_i(\mu) $ denote
the local operators generated by QCD and electroweak interactions.
$ C_i(\mu) $ stand for the Wilson
coefficient functions (c-numbers). 
The scale $ \mu $ separates the physics contributions in the ``short
distance'' contributions (corresponding to scales higher than $\mu $)
contained in $ C_i(\mu) $ and the ``long distance'' contributions
(scales lower than $ \mu $) contained in 
$\langle F \mid Q_i (\mu) \mid M \rangle $.
 By evolving the scale from $ \mu ={\cal O}(\mw) $ down to 
lower values of $ \mu $ one transforms the physics information 
at scales higher
than $ \mu $ from the hadronic matrix elements into $ C_i(\mu) $. Since no
information is lost this way the full amplitude cannot depend on $ \mu $.
This is the essence of the renormalization group equations which govern the
evolution $ (\mu -dependence) $ of $ C_i(\mu) $. This $ \mu $-dependence 
must be cancelled by the one present in $\langle  Q_i (\mu)\rangle $.
Generally this cancellation involves many
operators due to the operator mixing under renormalization.

It should  be stressed that the use of the renormalization group
is  necessary in order to sum up large logarithms
 $ \log \mw/\mu $ which appear for $ \mu= {\cal O}(1-2\gev) $.
 In the so-called leading
logarithmic approximation (LO) terms $ (\alpha_s\log \mw/\mu)^n $ are summed.
The next-to-leading logarithmic correction (NLO) to this result involves
summation of terms $ (\alpha_s)^n (\log \mw/\mu)^{n-1} $ and so on.
This hierarchic structure gives the renormalization group improved
perturbation theory.

I will not discuss here the technical details of the renormalization group
and of the calculation of
$C_i(\mu)$. They can be found in a recent review \cite{BBL} and in
the contributions of Martinelli and Nierste to these proceedings. 
Let me just list a few operators which play an important role
in the phenomenology of K decays. These are ($\alpha$ and $\beta$ are
 colour indices):

{\bf Current--Current:}
\begin{equation}\label{O1} 
Q_1 = (\bar s_{\alpha} u_{\beta})_{V-A}\;(\bar u_{\beta} d_{\alpha})_{V-A}
~~~~~~Q_2 = (\bar s u)_{V-A}\;(\bar u d)_{V-A} 
\end{equation}

{\bf QCD--Penguin  and Electroweak--Penguin:}
\begin{equation}\label{O2}
 Q_6 = (\bar s_{\alpha} d_{\beta})_{V-A}\sum_{q=u,d,s}
       (\bar q_{\beta} q_{\alpha})_{V+A} 
~~~~~ Q_8 = {3\over2}\;(\bar s_{\alpha} d_{\beta})_{V-A}\sum_{q=u,d,s}e_q
        (\bar q_{\beta} q_{\alpha})_{V+A}
\end{equation}

{\bf $\Delta S = 2 $ and $ \Delta B = 2 $ Operators:}
\begin{equation}\label{O7}
Q(\Delta S = 2)  = (\bar s d)_{V-A} (\bar s d)_{V-A}~~~~~
 Q(\Delta B = 2)  = (\bar b d)_{V-A} (\bar b d)_{V-A} 
\end{equation}

{\bf Semi--Leptonic Operators:}
\begin{equation}\label{9V}
Q_{9V}  = (\bar s d  )_{V-A} (\bar e e)_{V}~~~~~
Q_{10A}  = (\bar s d )_{V-A} (\bar e e)_{A}
\end{equation}
\begin{equation}\label{9NU}
Q_{\bar\nu\nu}  = (\bar s d )_{V-A} (\bar\nu\nu)_{V-A}~~~~~
Q_{\bar\mu\mu}  = (\bar s d )_{V-A} (\bar\mu\mu)_{V-A}
\end{equation}

\begin{figure}[hbt]
\vspace{0.10in}
\centerline{
\epsfysize=4in
\rotate[r]{
\epsffile{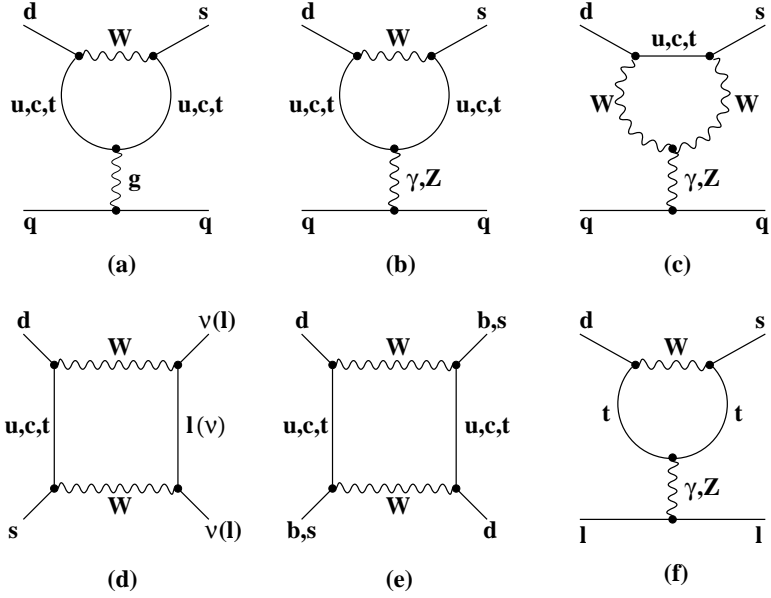}
}}
\vspace{0.08in}
\caption[]{
Typical Penguin and Box Diagrams.
\label{fig:fdia}}
\end{figure}

The rather formal expression for the decay amplitudes given in
(\ref{OPE}) can always be cast in the form \cite{PBE}:
\begin{equation}\label{PBEE}
A(M\to F)=\sum_i B_i {\rm V_{CKM}}^{i} \eta^{i}_{QCD} F_i(\mt,\mc)
\end{equation}
which is more useful for phenomenology. In writing (\ref{PBEE})
we have generalized (\ref{OPE}) to include several CKM factors.
$F_i(m_t,m_c)$, the Inami-Lim functions \cite{IL}, 
 result from the evaluation of loop diagrams with
internal top and charm exchanges (see fig. 3) and may also depend
solely on $\mt$ or $\mc$. In the case of new physics they  
depend  on masses of new particles such as charginos, stops, charged
Higgs scalars etc. 
The factors $\eta^{i}_{QCD}$ summarize
the QCD corrections which can be calculated by formal methods
discussed above. Finally $B_i$ stand for nonperturbative factors
related to the hadronic matrix elements of the contributing
operators: the main theoretical uncertainty in the whole enterprise.
In leptonic and semi-leptonic decays for which only the matrix elements
of weak currents are needed,  
the non-perturbative $B$-factors can fortunately be determined from
leading tree level decays reducing
or removing the non-perturbative uncertainty. In non-leptonic
decays this is generally not possible and we have to rely on
existing non-perturbative methods. A well known example of a
$B_i$-factor is the renormalization group invariant parameter 
$B_K$  defined by 
\begin{equation}\label{bk}
B_K=B_K(\mu)\left[\alpha_s(\mu)\right]^{-2/9}
\qquad
\langle \bar K^{0}\mid Q(\Delta S=2)\mid K^{0}\rangle=
\frac{8}{3} B_K(\mu)F_K^2 m_K^2
\end{equation}
For simplicity we did not show the NLO
correction in $B_K$.
$B_K$ plays an important role in the phenomenology of CP violation
in $K \to \pi\pi$. 

In order to achieve sufficient precision the Wilson coefficients
or equvalently the QCD factors $\eta^{i}_{QCD}\equiv\eta_i$ have 
to include both
the leading and the next-to-leading (NLO) corrections.
These corrections are known by now
for the most important and interesting decays and are 
reviewed in \cite{BBL}. 
We will discuss the impact of NLO calculations below. 

Let us recall why NLO calculations are important for the
phenomenology of weak decays:

\begin{itemize}
\item The NLO is first of all necessary to test the validity of
the renormalization group improved perturbation theory.
\item Without going to NLO the QCD scale $\Lambda_{\overline{MS}}$
extracted from various high energy processes cannot be used 
meaningfully in weak decays.
\item Due to the renormalization group invariance the physical
amplitudes do not depend on the scales $\mu$ present in $\alpha_s$
or in the running quark masses, in particular $\mt(\mu_t)$, 
$\mb(\mu_b)$ and $\mc(\mu_c)$. However
in perturbation theory this property is broken through the truncation
of the perturbative series. Consequently one finds sizable scale
ambiguities in the leading order, which can be reduced considerably
by going to NLO.
\item In several cases the central issue of the top quark mass dependence
is strictly a NLO effect.
\end{itemize}

Clearly in order to calculate the full amplitude in (\ref{PBEE}) or 
(\ref{OPE}) also the $B_i$ factors or
the matrix elements $\langle F \mid Q_i (\mu) \mid M \rangle$ have to
be evaluated. Since they involve long distance contributions one is
forced in this case to use non-perturbative methods.
Several non-perturbative methods have been discussed at this workshop.
Let me make only a few remarks here. The details can be found in the
corresponding contributions to these proceedings.

The lattice calculations have been discussed by Martinelli and Kilcup.
The progress in this field is rather slow but eventually this could
be the most powerful method to evaluate hadronic matrix elements.
Yet it seems that in the case of non-leptonic decays we have to
wait 5 to 15 years before calculation at the level of 10\% (statistic and
systematic) could be achieved. Some lattice results will be
given below.

Other methods have been reviewed by Donoghue in a very enjoyable talk 
and specifically discussed
by Bijnens, Bertolini, Fabbrichesi, Fajfer, Pich, Singer and Soldan. 
In particular
Bijnens stressed the great potential of DA${\Phi}$NE in testing chiral
perturbation theory. At this point I can only recommend to read his
contribution and to study "The second DA${\Phi}$NE Physics Handbook"
which is a great piece of work. 
Yet since chiral parturbation theory is based
mainly on symmetries it has its own limitations which are particularly
seen in the important work of Kambor, Missimer and Wyler \cite{KMW} where the
the general form of $\Delta S =1$ non-leptonic Lagrangian  at 
${\cal O}(p^4)$ has been worked out.
At this level there are too many unknown parameters (counter terms) which 
cannot be determined experimentally and one has to use some additional
dynamical methods like $1/N$ \cite{BBG,Cheng,Espriu} 
in order to have some predictions. 
Yet as stressed by Pich,
reviewing his work with de Rafael, Ecker and other spanish masters
(Bruno and Prades),
in other decays such as $K_S \to \gamma\gamma$, $K_L\to\pi^0\gamma\gamma$,
$K^+\to \pi^+\gamma\gamma$ (seen for the first time by BNL-E787) 
the situation is very different and 
predictions can be made easier. See also a recent work on 
$K^+\to \pi^+\gamma\gamma$  in \cite{Ambrosio}. 
One should hope that the status of non-leptonic
decays in the framework of chiral perturbation theory will improve in the 
DA${\Phi}$NE era.

Donoghue discussed also his interesting work with Gabbiani which uses
dispersion relations. Bertolini and Fabbrichesi on the other hand 
presented their extensive 
calculations of $\Delta I=1/2$ rule and of $\epe$ in the
framework of the Chiral Quark Model. This model has potential
uncertainties related to the value of the gluon condensate, which
plays an important role in enhancing the $\Delta I=1/2$ transitions.
This "gluon condensate effect"
has been already identified by Pich and de Rafael in their work of 1991
\cite{PRAF} 
and is another expression of  the $\Delta I=1/2$ enhancement through
quadratic cut-off dependence in the 1/N approach of Bardeen, G\`erard
and myself \cite{BBG} which we discussed almost ten years ago. Yet it is
important to see such effects in different settings. Let us hope
that one day this rule will be understood at a fully quantitative level
and that the relation of the approaches mentioned above to the
diquark approach of Stech and Neubert \cite{SN} will be  clarified.
Some insight here can be gained from the work of Jamin and Pich \cite{JP}.

Clearly the list of non-perturbative methods has to include QCD
sum rules \cite{QCDS} which play a substantial role in estimating
non-perturbative parameters. Finally I would like to mention the 
interesting work on the extended Nambu-Jona-Lasinio model by 
Bijnens, Bruno and de Rafael \cite{BBR}. Other non-perturbative 
strategies are reviewed in \cite{Belkov}.

 Needless 
to say all the non-perturbative methods listed above
have  at present considerable limitations and only time will
show whether the situation can be considerably improved.
 Consequently the dominant theoretical
uncertainties in the decay amplitudes reside in
the matrix elements of $Q_i$ or the corresponding $B_i$ factors.

After these general remarks let us move to the presentation of some
results obtained in 
this framework and its confrontation with the experimental data.

\section{Standard Analysis}

\subsection{Basic Formulae}

At present there is still a rather limited knowledge of the shape of 
the unitarity triangle. The standard analysis using the available
experimental and theoretical information proceeds essentially in five
steps:

{\bf Step 1:}

{}From  $b\to c$ transition in inclusive and exclusive B meson decays
one finds $\vcb$ and consequently the scale of UT:
\begin{equation}
\vcb\quad =>\quad\lambda \vcb= \lambda^3 A
\end{equation}

{\bf Step 2:}

{}From  $b\to u$ transition in inclusive and exclusive B meson decays
one finds $\vub$ and consequently the side $CA=R_b$ of UT:
\begin{equation}
\vub \quad=> \quad R_b=4.44 \cdot \left| \frac{V_{ub}}{V_{cb}} \right|
\end{equation}
\newpage
{\bf Step 3:}

{}From the observed indirect CP violation in $K \to \pi\pi$ described
experimentally by the parameter $\varepsilon_K$ and theoretically
by the imaginary part of the relevant box diagram in fig. 3 one 
derives the constraint:
\begin{equation}\label{100}
\bar\eta \left[(1-\bar\varrho) A^2 \eta_2 S(x_t)
+ P_0(\varepsilon) \right] A^2 B_K = 0.226
\qquad
S(x_t)=0.784 \cdot x_t^{0.76}
\end{equation}
where
\begin{equation}\label{102}
P_0(\varepsilon) = 
\left[ \eta_3 S(x_c,x_t) - \eta_1 x_c \right] \frac{1}{\lambda^4}
\qquad
x_t=\frac{\mt^2}{\mw^2}
\end{equation}
Equation (\ref{100}) specifies 
a hyperbola in the $(\bar \varrho, \bar\eta)$
plane. Here $B_K$
is the non-perturbative parameter defined in (\ref{bk}) and $\eta_2$
is the QCD factor in the box diagrams with two top quark exchanges.
Finally $P_0(\varepsilon)=0.31\pm0.02$ summarizes the contributions
of box diagrams with two charm quark exchanges and the mixed 
charm-top exchanges. $P_0(\varepsilon)$ depends very weakly on $m_t$ and
its range given above corresponds to $155~GeV \le m_t \le 185~GeV$.
The NLO values of the QCD factors $\eta_1$ , $\eta_2$ and $\eta_3$ 
are given as follows \cite{HNa,BJW,HNb}:
\begin{equation}
\eta_1=1.38\pm 0.20\qquad
\eta_2=0.57\pm 0.01\qquad
  \eta_3=0.47\pm0.04
\end{equation}

The quoted errors reflect the remaining theoretical uncertainties due to
$\Lambda_{\overline{MS}}$ and the quark
masses. The references to the leading order calculations can be found in
\cite{BBL}. The factor $\eta_1$ plays only a minor role in the analysis of
$\varepsilon_K$ but its enhanced value through NLO corrections \cite{HNa}
is essential for the $K_L-K_S$ mass difference.

Concerning the parameter $B_K$, the most recent analyses
using the lattice methods summarized by Kilcup here and recently by 
Flynn \cite{Flynn} give
$B_K=0.90\pm 0.06$. The $1/N$ approach
 of \cite{BBG0}  gives  $B_K=0.70\pm 0.10$. A recent confirmation of this
result in a somewhat modified framework has been presented by 
Bijnens and Prades \cite{Bijnens}  who gave plausible arguments for 
the difference between this result
for $B_K$ and the lower values obtained by using the QCD Hadronic Duality
approach \cite{Prades} ($B_K=0.39\pm 0.10$) or using the $SU(3)$ symmetry and
 PCAC
($B_K=1/3$) \cite{Donoghue}. For $\vcb=0.040$ and $\vub=0.08$ such 
low values for
$B_K$ require $\mt>200~GeV$ in order to explain the experimental
value of $\varepsilon_K$ \cite{AB,BLO,HNb}. The QCD sum rule results are in
the ball park of $B_K=0.60$ \cite{DENA}. In our numerical analysis presented 
below we will use $B_K=0.75\pm 0.15$ (see table 1).

{\bf Step 4:}

{}From the observed $B^0_d-\bar B^0_d$ mixing described experimentally 
by the mass difference $(\Delta M)_d$ or by the
mixing parameter $x_d=\Delta M/\Gamma_B$ 
and theoretically by the relevant box diagram of fig. 3
the side $BA=R_t$ of the UT can be determined:
\begin{equation}\label{106}
 R_t = 1.0 \cdot
\left[\frac{|V_{td}|}{8.7\cdot 10^{-3}} \right] 
\left[ \frac{0.040}{\vcb} \right]
\end{equation}
with
\begin{equation}\label{VT}
\vtd=
8.7\cdot 10^{-3}\left[ 
\frac{200\mev}{\sqrt{B_{B_d}}F_{B_d}}\right]
\left[\frac{170~GeV}{\mtb(\mt)} \right]^{0.76} 
\left[\frac{(\Delta M)_d}{0.45/ps} \right ]^{0.5} 
\sqrt{\frac{0.55}{\eta_B}}  
\end{equation}

Here $\eta_B$ is the QCD factor analogous to $\eta_2$ and given
by $\eta_B=0.55\pm0.01$ \cite{BJW}. Next 
$F_{B_d}$ is the B-meson decay constant and $B_{B_d}$ 
denotes a non-perturbative
parameter analogous to $B_K$. 
 
There is a vast literature on the lattice calculations of $F_{B_d}$ 
and $B_{B_d}$.
The most recent world averages are given by Flynn \cite{Flynn}:
\begin{equation}
F_{B_d}\sqrt{B_{B_d}}=175\pm 25~MeV \qquad
B_{B_d}=1.31\pm 0.03
\end{equation}
This result for $F_{B_d}$ is compatible with the results obtained using
QCD sum rules \cite{QCDSF}. An interesting upper bound 
$F_{B_d}<195~MeV$ using QCD dispersion relations can be found in
\cite{BGL}. In our numerical analysis we will use
$F_{B_d}\sqrt{B_{B_d}}=200\pm 40~MeV$. The experimental situation on
$(\Delta M)_d$ has been recently summarized by Gibbons \cite{Gibbons}
 and is given in table 1. For $\tau(B_d)=1.55~ps$ one has then 
$x_d= 0.72\pm 0.03$.

{\bf Step 5:}

{}The measurement of $B^0_s-\bar B^0_s$ mixing parametrized by $(\Delta M)_s$
together with $(\Delta M)_d$  allows to determine $R_t$ in a different
way. Setting $(\Delta M)^{max}_d= 0.482/ps$ and 
$|V_{ts}/V_{cb}|^{max}=0.993$ (see table 1) I find 
a useful formula:
\begin{equation}\label{107b}
(R_t)_{max} = 1.0 \cdot \xi \sqrt{\frac{10.2/ps}{(\Delta M)_s}}
\qquad
\xi = 
\frac{F_{B_s} \sqrt{B_{B_s}}}{F_{B_d} \sqrt{B_{B_d}}} 
\end{equation}
where $\xi=1$ in the  SU(3)--flavour limit.
Note that $\mt$ and $|V_{cb}|$ dependences have been eliminated this way
 and that $\xi$ should in principle 
contain much smaller theoretical
uncertainties than the hadronic matrix elements in $(\Delta M)_d$ and 
$(\Delta M)_s$ separately.  

The most recent values relevant for (\ref{107b}) are:
\begin{equation}\label{107c}
(\Delta M)_s > 9.2/ ps
\qquad
\xi=1.15\pm 0.05
\end{equation}
The first number is the improved lower bound quoted in \cite{Gibbons}
based in particular on ALEPH and DELPHI results.
The second number comes from quenched lattice calculations summarized
by Flynn in \cite{Flynn}.
A similar result has been obtained using QCD sum rules \cite{NAR}.
On the other hand another recent quenched lattice calculation \cite{Soni}
not included in (\ref{107c}) finds
$\xi\approx 1.3 $. Moreover one expects that unquenching will increase
the value of $\xi$ in (\ref{107c}) by roughly 10\% so that values as
high as $\xi=1.25-1.30$ are certainly possible even from Flynn's point of
view. For such high values of $\xi$
the lower bound on $(\Delta M)_s$ in (\ref{107c}) implies $R_t\le 1.37$
which as we will see is similar to the bound obtained on the basis of
the first four steps alone. On the other hand for $\xi=1.15$ one finds
$R_t \le 1.21 $ which puts an additional constraint on the unitarity
triangle cutting lower values of $\bar\varrho$ and higher values 
of $|V_{td}|$. In
view of remaining large uncertainties in $\xi$ we will not use the
constraint from $(\Delta M)_s$ below.

\subsection{Numerical Results}
\subsubsection{Input Parameters}
 The input parameters needed to perform the
standard analysis are given in table 1. The details on the chosen
ranges of $\left| V_{cb} \right|$ and $\left| V_{ub}/V_{cb} \right|$
can be found in \cite{Gibbons}.
Clearly during the last two years there has been a considerable progress
done by experimentalists and theorists in the extraction of
$\left| V_{cb} \right|$ from exclusive and inclusive decays. In
particular I would like to mention important papers by
Shifman, Uraltsev and Vainshtein \cite{SUV},
Neubert \cite{Neubert} and 
Ball, Benecke and Braun \cite{Braun}
on the basis of which one is entitled to use the value given in table 1.
In the case of $\left| {V_{ub}}/{V_{cb}} \right|$
the situation is much worse but progress in the next few years is to be
expected in particular due to new information coming from
exclusive decays \cite{CLEOU,Gibbons} and the inclusive semileptonic
$b\to u$ rate \cite{SUV,Braun,URAL}.

Next it is important to stress that the discovery of the top quark 
by CDF and D0 and its impressive mass measurement summarized recently by
Tipton \cite{Tipton} had an important impact on
the field of rare decays and CP violation reducing considerably one
potential uncertainty. At this point it should be recalled that 
the parameter $\mt$, the top quark mass, used in weak decays is not
equal to the one used in the electroweak precision studies at LEP,
SLD and FNAL. In the latter investigations the so-called pole mass is used,
whereas in all the NLO calculations   $\mt$ refers
to the running current top quark mass normalized at $\mu=\mt$:
$\mtb(\mt)$. 
For $\mt={\cal O}(170~GeV)$, $\mtb(\mt)$ is typically
by $8~GeV$ smaller than $\mt^{Pole}$. This difference matters already
because the most recent pole mass value has a very small error,
 $175\pm 6~GeV$ \cite{Tipton}, implying
$167\pm 6~GeV$ for $\mtb(\mt)$.
 In this review we will often denote this mass by $\mt$.

\begin{table}[thb]
\begin{center}
\begin{tabular}{|c|c|c|}
\hline
{\bf Quantity} & {\bf Central} & {\bf Error} \\
\hline
$|V_{cb}|$ & 0.040 & $\pm 0.003$ \\
$|V_{ub}/V_{cb}|$ & 0.080 & $\pm 0.020$ \\
$B_K$ & 0.75 & $\pm 0.15$ \\
$\sqrt{B_d} F_{B_{d}}$ & $200\mev$ & $\pm 40\mev$ \\
$\sqrt{B_s} F_{B_{s}}$ & $240\mev$ & $\pm 40\mev$ \\
$\mt$ & $167\gev$ & $\pm 6\gev$ \\
$(\Delta M)_d$ & $0.464~ps^{-1}$ & $\pm 0.018~ps^{-1}$ \\
$\Lms^{(4)}$ & $325\mev$ & $\pm 80\mev$ \\
\hline
\end{tabular}
\caption[]{Collection of input parameters.\label{tab:inputparams}}
\end{center}
\end{table}

\subsubsection{\bf $\left| V_{ub}/V_{cb} \right|$,
$\left| V_{cb} \right|$ and $\varepsilon_K$}

The values for $\left| V_{ub}/V_{cb} \right|$ 
and $\left| V_{cb} \right|$ in table 1 
are not correlated with
each other. On the other hand such a correlation is present in
the analysis of the CP violating parameter $\varepsilon_K$ which
is roughly proportional to the fourth power of $\left| V_{cb}\right|$
and linear in $\left|V_{ub}/V_{cb} \right|$. It follows
that not all values in table 1 are simultaneously
consistent with the observed value of $\varepsilon_K$. This is indirectly
seen in \cite{AB} and has been more explicitly emphasized last year by
Herrlich and Nierste \cite{HNb} and in \cite{BBL}. 
Updating and rewriting the analytic
lower bound on $\mt$ from $\varepsilon_K$ \cite{AB} one finds \cite{BBL}

\begin{equation}
\left| \frac{V_{ub}}{V_{cb}} \right|_{min}=
\frac{0.225}{B_K A^2(2 x_t^{0.76}A^2+1.4)}
\end{equation}

\begin{figure}[hbt]
\vspace{0.10in}
\centerline{
\epsfysize=4.5in
\rotate[r]{
\epsffile{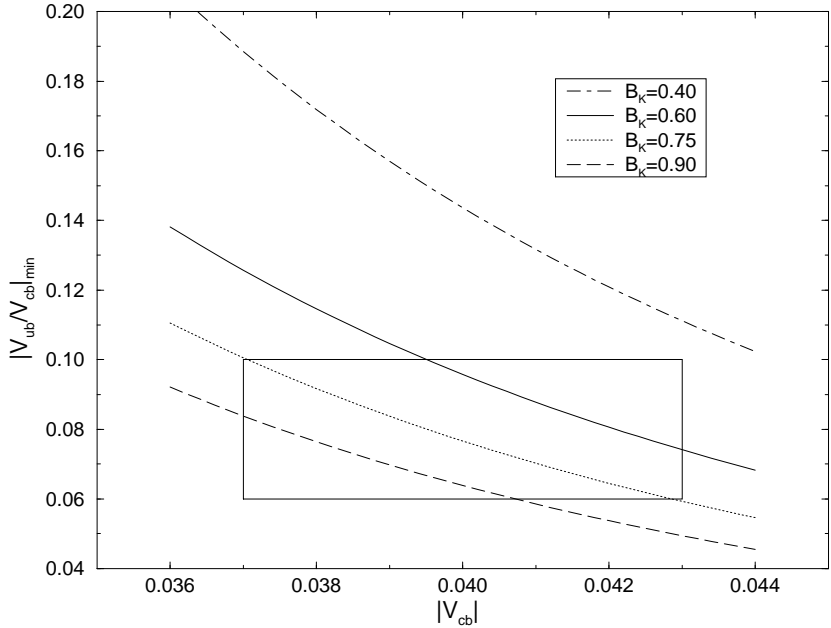}
}}
\vspace{0.08in}
\caption[]{
Lower bound on $\vub$ from $\varepsilon_K$.
\label{fig:bound}}
\end{figure}

This bound is shown as a function of $\mid V_{cb} \mid$ for different
values of $B_K$ and $\mt=173~GeV$ in fig.4. We observe that simultaneously
small values of $\left| V_{ub}/V_{cb} \right|$ and $\left| V_{cb} \right|$
although still consistent with the ones given in table 1, are not allowed
by the size of the indirect CP violation observed in $K \to \pi\pi$.

\subsubsection{Output of a Standard Analysis}
The output of the standard analysis depends to some extent on the
error analysis. This should be always remembered in view of the fact
that different authors use different procedures. In order to illustrate
this  I show in table 2 the results for various quantities of interest
using two types of the error analyses:

\begin{itemize}
\item
Scanning: Both the experimentally measured numbers and the theoretical input
parameters are scanned independently within the errors given in
table~\ref{tab:inputparams}. 
\item
Gaussian: The experimentally measured numbers and the theoretical input 
parameters are used with Gaussian errors.
\end{itemize}
Clearly the "scanning" method is a bit conservative. On the other
hand using Gaussian distributions for theoretical input parameters
can certainly be questioned. One could instead use flat distributions
(with a width of $2\sigma$) for the theoretical input parameters
as done in \cite{ciuchini:95}. The latter method gives however
similar results to the "Gaussian method". Personally I think that
at present the conservative "scanning" method should be preferred.
In the future however when data and theory improve, it would be useful to  
find a less conservative estimate which most probably will give errors
somewhere inbetween these two error estimates. 
The analysis discussed here has been done in collaboration with Matthias 
Jamin and Markus Lautenbacher. More details and more results can be found in 
\cite{BJL96b}.

\begin{table}
\begin{center}
\begin{tabular}{|c||c||c|}\hline
{\bf Quantity} & {\bf Scanning} & {\bf Gaussian} \\ \hline
$\mid V_{td}\mid/10^{-3}$ &$6.9 - 11.3$ &$ 8.6\pm 1.1$ \\ \hline
$\mid V_{ts}/V_{cb}\mid$ &$0.959 - 0.993$ &$0.976\pm 0.010$  \\ \hline
$\mid V_{td}/V_{ts}\mid$ &$0.16 - 0.31$ &$0.213\pm 0.034$  \\ \hline
$\sin(2\beta)$ &$0.36 - 0.80$ &$ 0.66\pm0.13 $ \\ \hline
$\sin(2\alpha)$ &$-0.76 - 1.0$ &$ 0.11\pm 0.55 $ \\ \hline
$\sin(\gamma)$ &$0.66 - 1.0 $ &$ 0.88\pm0.10 $ \\ \hline
$\IM \lambda_t/10^{-4}$ &$0.86 - 1.71 $ &$ 1.29\pm 0.22 $ \\ \hline
$(\Delta M)_s ps$ &$ 8.0 - 25.4$ &$ 15.2 \pm 5.5 $ \\ \hline
\end{tabular}
\caption[]{Output of the Standard Analysis. 
 $\lambda_t=V^*_{ts} V_{td}$.\label{TAB2}}
\end{center}
\end{table}

\begin{figure}[hbt]
\vspace{0.10in}
\centerline{
\epsfysize=6in
\rotate[r]{
\epsffile{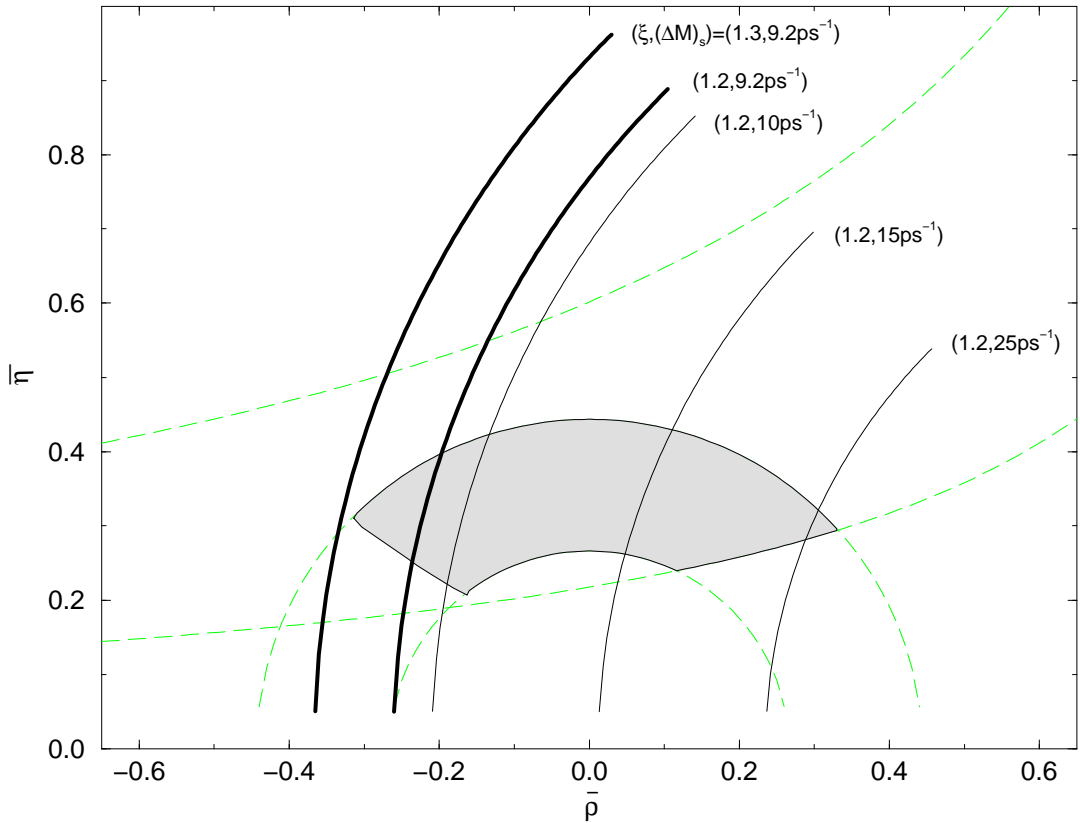}
}}
\vspace{0.08in}
\caption[]{
Unitarity Triangle 1996.
\label{fig:utdata}}
\end{figure}

In fig. 5  we show the range for the upper
corner A of the UT. The solid thin lines correspond to $R_t^{max}$ from 
(\ref{107b})
using $\xi=1.20$ and $(\Delta M)_s=10/ps,~15/ps$ and $25/ps$, respectively.
The allowed region has a typical "banana" shape which can be found
in many other analyses \cite{BLO,ciuchini:95,HNb,ALUT,PP,PW}. The size of
the banana and its position depends on the assumed input
parameters and on the error analysis which varies from paper
to paper. The results in fig. 5 correspond to a simple independent 
scanning of all parameters within one standard deviation.
Effectively such an approach is more conservative than using
Gaussian distributions as done in some papers quoted above.
We show also the impact of the experimental bound $(\Delta M)_s>9.2/ps$
with $\xi=1.20$ and the corresponding bound for $\xi=1.30$. In view
of the remaining uncertainty in $\xi$, in particular due to quenching,
 this bound has not 
been used in
obtaining the results in table 2. It is evident however that $B^0_s-\bar
B^0_s$ mixing will have a considerable impact on the unitarity triangle
when the value of $\xi$ will be better know and the data improves.
This is very desirable because as seen in fig. 5 our knowledge of
the unitarity triangle is still rather poor. Similarly the uncertainty
in the predicted value of $(\Delta M)_s$ using $\sqrt{B_s} F_{B_{s}}$
of table 1 
is large with central values
around $15/ps$. 
\section{ $\varepsilon'/\varepsilon$}

The measurement of $\varepsilon'/\varepsilon$ at the $10^{-4}$ level
remains as one of the important targets of contemporary particle
physics. A non-vanishing value of this ratio would give the first
signal for the direct CP violation ruling out the superweak models.
The experimental situation on Re($\varepsilon'/\varepsilon$) is unclear
at present. While the result of NA31 collaboration at CERN with
Re$(\varepsilon'/\varepsilon) = (23 \pm 7)\cdot 10^{-4}$ \cite{barr:93}
clearly indicates direct CP violation, the value of E731 at Fermilab,
Re$(\varepsilon'/\varepsilon) = (7.4 \pm 5.9)\cdot 10^{-4}$
\cite{gibbons:93}, is compatible with superweak theories
\cite{WO1} in which $\varepsilon'/\varepsilon = 0$.
 Hopefully, in about two years the experimental situation concerning
$\varepsilon'/\varepsilon$ will be clarified through the improved
measurements by the two collaborations at the $10^{-4}$ level and by
the KLOE experiment at  DA$\Phi$NE.

There is no question about  that the direct CP violation is present in
the standard model. Yet accidentally it could turn out that it will be
difficult to see it in $K \to \pi\pi$ decays.  Indeed in the standard
model $\varepsilon'/\varepsilon $ is governed by QCD penguins and
electroweak (EW) penguins. In spite of being suppressed by
$\alpha/\alpha_s$ relative to QCD penguin contributions, the
electroweak penguin contributions have to be included because of the
additional enhancement factor ${\rm Re}A_0/{\rm Re}A_2=22$ relative
to QCD penguins. With increasing $\mt$ the EW penguins become
increasingly important \cite{flynn:89,buchallaetal:90}, and entering
$\varepsilon'/\varepsilon$ with the opposite sign to QCD penguins
suppress this ratio for large $\mt$. For $\mt\approx 200\,\gev$ the ratio
can even be zero \cite{buchallaetal:90}.  Because of this strong
cancellation between two dominant contributions and due to uncertainties
related to hadronic matrix elements of the relevant local operators, a
precise prediction of $\varepsilon'/\varepsilon$ is not possible at
present.

In spite of all these difficulties, a considerable progress has been
made in this decade to calculate $\varepsilon'/\varepsilon$. First of
all the complete next-to-leading order (NLO) effective hamiltonians for
$\Delta S=1$ \cite{burasetal:92a,burasetal:92d,ciuchini:93}, $\Delta S=2$
\cite{BJW,HNa,HNb} and $\Delta B=2$ \cite{BJW} are now available so that a
complete NLO analysis of $\varepsilon'/\varepsilon$ including
constraints from the observed indirect CP violation ($\varepsilon_K$)
and the $B^0_d-\bar B^0_d$ mixing ($(\Delta M)_d$) is possible. The improved
determination of the $V_{ub}$ and $V_{cb}$ elements of the CKM matrix
\cite{Gibbons}, and in particular the determination of the top quark mass
$\mt$ \cite{Tipton} had of course also an important impact on
$\varepsilon'/\varepsilon$. The main remaining theoretical
uncertainties in this ratio are then the poorly known hadronic matrix
elements of the relevant QCD penguin and electroweak penguin operators
represented by  two important B-factors
($B_6=$ the dominant QCD penguin $Q_6$ and $B_8=$ the dominant electroweak
 penguin $Q_8$), the values of the ${\rm V_{CKM}}$ factors
 and as stressed in \cite{burasetal:92d} the value of $\ms$ and
$\Lambda_{\overline{MS}}$.

 An analytic formula for
$\varepsilon'/\varepsilon$ which exhibits all these uncertainties
can be found in \cite{BLAU,BJL96a}. 
A very simplified version of this formula is given as follows
\begin{equation}\label{7e}
\frac{\varepsilon'}{\varepsilon}=11\cdot 10^{-4}\left[ 
\frac{\eta\lambda^5 A^2}{1.3\cdot 10^{-4}}\right]
\left[\frac{140~MeV}{ m_s(2~GeV)} \right]^2 
\left[\frac{\Lms^{(4)}}{300~MeV} \right]^{0.8} 
[B_6-Z(x_t)B_8]
\end{equation} 
where $Z(x_t)\approx 0.18 (\mt/\mw)^{1.86}$ and equals unity for 
$\mt\approx 200~GeV$. This simplified formula should not be used
for any serious numerical analysis.

Concerning the values of $B_6$ and $B_8$ one has $B_6=B_8=1$ in the 
vacuum insertion estimate of the hadronic matrix elements in question.
The same result is found in the large $N$ limit
\cite{bardeen:87,burasgerard:87}. Also lattice calculations give
similar results: $B_6=1.0 \pm 0.2$ \cite{kilcup:91,sharpe:91} and $B_8
= 1.0 \pm 0.2$ \cite{kilcup:91,sharpe:91,bernardsoni:89,francoetal:89},
$B_8= 0.81(1)$ \cite{gupta2:96}. These are the values used in
\cite{burasetal:92d,ciuchini:95,BBL,BJL96a}.
In the chiral quark model one finds \cite{bertolinietal:95}: $B_6=1.0\pm0.4$,
$B_8=2.2\pm1.5$ and generally $B_8>B_6$. On the other hand the Dortmund group
\cite{heinrichetal:92, paschos:96} advocates  $B_6>B_8$.
From \cite{paschos:96} $B_6=1.3$ and $B_8=0.7$ can be extracted.
As discussed by Soldan at this workshop, a new Dortmund calculation is
in progress. Other technical details can be found in the talks of
Bertolini and Martinelli.

At this point it seems appropriate to summarize the present status of 
the value of the strange quark mass. The most recent results of QCD sum
rule (QCDSR) calculations \cite{jaminmuenz:95,chetyrkinetal:95,narison:95}
obtained at $\mu=1\gev$ correspond to $\ms(2\gev)=145\pm20\mev$.
The lattice calculation of \cite{alltonetal:94} finds
$\ms(2~GeV)=128\pm18~MeV$, 
in rather good agreement with the QCDSR result. This summer a new lattice
result has been presented by Gupta and Bhattacharya \cite{gupta:96}. They
find $\ms(2\gev)=90\pm20\mev$ which is on the low side of all strange mass 
determinations. Moreover they find that unquenching lowers further
the values of $\ms(2\gev)$ to $70\pm15\mev$. 
Similar results are found by the FNAL 
group \cite{FNALL}.
The situation with the strange quark mass is therefore unclear at present
and hopefully will be clarified soon. 

It should also be remarked that the decomposition of the relevant hadronic
matrix elements of penguin operators into a product of $B_i$ factors times
$1/\ms^2$ although useful in the $1/N$ approach is unnecessary in a brute
force method like the lattice approach. It is to be expected that 
future lattice calculations will directly give the relevant hadronic 
matrix elements and the issue of $\ms$ in connection withe $\epe$ will
disappear.

The most recent analysis of \cite{BJL96a} using input parameters of table 1,
$B_6=1.0\pm 0.2$, $B_8=1.0\pm 0.2$ and 
$m_s(2~GeV)=129\pm17\mev$ finds
\begin{equation}
-1.2 \cdot 10^{-4} \le \epe \le 16.0 \cdot 10^{-4}
\label{eq:eperangenew}
\end{equation}
and
\begin{equation}
\epe= ( 3.6\pm 3.4) \cdot 10^{-4}
\label{eq:eperangefinal}
\end{equation}
for the "scanning" method and the "gaussian" method respectively.

The result in (\ref{eq:eperangefinal}) agrees rather well with
the 1995 analysis of the Rome group \cite{ciuchini:95} which finds
 $\varepsilon'/\varepsilon=(3.1\pm 2.5)\cdot 10^{-4}$.
On the other hand the range in (\ref{eq:eperangenew}) shows that for
particular choices of the input parameters, values for $\epe$ as high as
$16\cdot 10^{-4}$ cannot be excluded at present. Such high values are
found if simultaneously  $\vub=0.10$, $B_6=1.2$, $B_8=0.8$, $B_K=0.6$,
$\ms(2\gev)=110 MeV$, $\Lms^{(4)}=405\mev$ and low values of $\mt$ still
consistent with the $\varepsilon_K$ and the observed $B_d^0-\bar B_d^0$ 
mixing
are chosen. It is however evident from  the comparision of
(\ref{eq:eperangenew}) and (\ref{eq:eperangefinal})  that such 
high values of $\epe$ and generally values above $10^{-3}$ 
are very improbable.

The authors of \cite{bertolinietal:95} calculating the $B_i$ factors
in the chiral quark model find using the scanning method
 a rather large range $-50 \cdot 10^{-4}\le \epe \le 14 \cdot 10^{-4}$.
 In particular they find in contrast to
\cite{burasetal:92d,ciuchini:95,BBL,BJL96a} 
that negative values for $\epe$ as large as $-5\cdot
10^{-3}$ are possible. 
The Dortmund group
\cite{heinrichetal:92} advocating on the other hand $B_6>B_8$ finds
$\epe=(9.9\pm 4.1)\cdot 10^{-4}$ for $m_s(2~GeV)=130~MeV$ \cite{paschos:96}.
From the point of view of the analyses in 
\cite{ciuchini:95,BJL96a} 
such high values of $\epe$ for $\ms(2~GeV)={\cal O}(130~MeV)$
are rather improbable within the standard model.

The situation with $\epe$ in the standard model may however change
 if the value for $m_s$ is as low as found in \cite{gupta:96}.
Using 
$\ms(2\gev)=85\pm17\mev$ one finds \cite{BJL96a}
\begin{equation}
0 \le \epe \le 43.0 \cdot 10^{-4}
\label{eq:eperangenewa}
\end{equation}
and
\begin{equation}
\epe= ( 10.4\pm 8.3) \cdot 10^{-4}
\label{eq:eperangefinala}
\end{equation}
for the "scanning" method and the "gaussian" method respectively.
We observe that the "gaussian" result agrees well with the E731
value and
as stressed in \cite{BJL96a} the decrease of $\ms$
 with $\ms(2\gev)\geq 85~MeV$ alone is insufficient to bring 
the standard model to agree with
the NA31 result. However for $B_6>B_8$, sufficiently large values of
$|V_{ub}/V_{cb}|$ and $\Lms$ and small values of $\ms$, the values
of $\epe$ in the standard model can be as large as $(2-4)\cdot 10^{-3}$
and consistent with the NA31 result.

Let us hope that the future experimental and theoretical results will
be sufficiently accurate to be able to see whether $\epe\not=0$ and
whether the standard model agrees with the data. In any case the
coming years should be very exciting. 

\section{Rare K Decays}
\subsection{The Decay $K_L\to\pi^0 \lowercase{e}^+\lowercase{e}^-$}

Let us next move on to discuss the rare decay $K_L\to\pi^0e^+e^-$.
Whereas in $K\to\pi\pi$ decays the CP violating contribution is
only a tiny part of the full amplitude and the direct CP violation
as we have just seen is expected to be at least by three orders of
magnitude smaller than the indirect CP violation, the corresponding
hierarchies are very different for $K_L\to\pi^0e^+e^-$. At lowest
order in electroweak interactions (one-loop photon penguin,
$Z^0$-penguin and W-box diagrams), this decay takes place only if
CP symmetry is violated. The CP conserving contribution to the
amplitude comes from a two photon exchange, which although of higher
order in $\alpha$ could in principle be sizable. 
The CP
violating part can again be divided into a direct and an indirect one.
The latter is given by the $K_S\to\pi^0e^+e^-$ amplitude times the CP
violating parameter $\varepsilon_K$.

Now as reviewed by Pich at this workshop, out of these three contributions 
only the the directly CP violating contribution can be calculated reliably.
The other two contributions are unfortunately very uncertain at present
and the following ranges can be found in the literature:
\begin{equation}
Br(K_L \to \pi^0 e^+ e^-)_{cons}\approx\left\{ \begin{array}{ll}
(0.3-1.8)\cdot 10^{-12} & \hbox{\cite{cohenetal:93}} \\
     4.0 \cdot 10^{-12} & \hbox{\cite{heiligerseghal:93}} \\
(5 \pm 5)\cdot 10^{-12} & \hbox{\cite{donoghuegabbiani:95}}
\end{array} \right.
\label{eq:BKLtheo}
\end{equation}
and \cite{eckeretal:88,brunoprades:93,heiligerseghal:93,donoghuegabbiani:95}
\begin{equation}
Br(K_L \to \pi^0 e^+ e^-)_{indir}=(1.-5.)\cdot 10^{-12} 
\label{eq:BKLindir2}
\end{equation}

In what follows, we will concentrate on the directly CP violating
contribution. 
There are
practically no theoretical uncertainties here 
because the relevant matrix element 
$\langle\pi^0|(\bar sd)_{V-A}| K_L\rangle$ can be
extracted using isospin symmetry from the well measured decay
$K^+\to\pi^0e^+\nu$. Calculating the relevant box and electroweak
penguin diagrams and including LO \cite{GIL2} and NLO QCD \cite{BLMM} 
corrections one finds
approximately:
\begin{equation}\label{KT}
Br(K_L \to \pi^0 e^+ e^-)_{dir}=
4.4\cdot 10^{-12}\left [ 
\frac{\eta}{0.37}\right ]^2
\left [\frac{|V_{cb}|}{0.040} \right ]^4 
\left [\frac{(\mtb(\mt)}{170\gev} \right ]^2 
\end{equation}
Scanning the input parameters of table 1 one finds \cite{BJL96b}
\begin{equation}\label{12}
Br(K_L\to\pi^0 e^+ e^-)_{dir}=(4.5 \pm 2.6)\cdot 10^{-12} 
\end{equation}
where the error comes dominantly from the uncertainties in the CKM
parameters. Thus the directly CP violating contribution is comparable
to the other two contributions. It is however possible that the direct
CP violation dominates in this decay which is of course very exciting.
In order to see whether this is indeed the case  improved estimates of the
other two contributions are necessary.

A much better assessment of the importance of the
indirect CP violation in $K_L\to\pi^0e^+e^-$ will become possible after
a measurement of $Br(K_S\to\pi^0e^+e^-)$.  Bounding the latter
branching ratio below $ 1 \cdot 10^{-9}$ or $ 1 \cdot 10^{-10}$ would
bound the indirect CP contribution below $ 3 \cdot 10^{-12}$ and
 $ 3 \cdot 10^{-13}$ respectively. The present bounds: 
$ 1.1 \cdot 10^{-6}$ (NA31) and  $ 3.9 \cdot 10^{-7}$ (E621) are still 
too weak. On the other hand KLOE at DA${\Phi}$NE
could make an important contribution here.  

The present experimental bounds
\begin{equation}
Br(K_L\to\pi^0 e^+ e^-) \leq\left\{ \begin{array}{ll}
4.3 \cdot 10^{-9} & \cite{harris} \\
5.5 \cdot 10^{-9} & \cite{ohl} \end{array} \right.
\end{equation}
are still by three orders of magnitude away from the theoretical
expectations in the Standard Model. Yet the prospects of getting the
required sensitivity of order $10^{-11}$--$10^{-12}$ by 1999 are
encouraging \cite{CPRARE}. More details on this interesting decay can
be found in the original papers and in the talk by Pich.

\subsection{$K_L\to\pi^0\nu\bar\nu$ and $K^+\to\pi^+\nu\bar\nu$}
$K_L\to\pi^0\nu\bar\nu$ and $K^+\to\pi^+\nu\bar\nu$ are the theoretically
cleanest decays in the field of rare K-decays. 
$K_L\to\pi^0\nu\bar\nu$ is 
dominated by short distance loop diagrams (Z-penguins and box diagrams)
involving the top quark.  $K^+\to\pi^+\nu\bar\nu$ receives
additional sizable contributions from internal charm exchanges.
The great virtue of $K_L\to\pi^0\nu\bar\nu$ is that it proceeds
almost exclusively through direct CP violation \cite{Littenberg} 
and as such is the
cleanest decay to measure this important phenomenon. It also offers
a clean determination of the Wolfenstein parameter $\eta$ and in particular
as we will stress in section 6 offers the cleanest measurement
of $\IM\lambda_t= \IM V^*_{ts} V_{td}$ 
which governs all  CP violating  K-decays. 
$K^+\to\pi^+\nu\bar\nu$ is CP conserving and offers a clean 
determination of $|V_{td}|$. Due to the presence of the charm
contribution and the related $m_c$ dependence it has a small
scale uncertainty absent in $K_L\to\pi^0\nu\bar\nu$.

The next-to-leading QCD corrections to both decays
have been calculated in a series of papers by Buchalla and
myself \cite{BB13}. These calculations
considerably reduced the theoretical uncertainty
due to the choice of the renormalization scales present in the
leading order expressions \cite{DDG}, in particular in the charm contribution
to $K^+\to\pi^+\nu\bar\nu$. Since the relevant hadronic matrix
elements of the weak currents entering $K\to \pi\nu\bar\nu$
can be related using isospin symmetry to the leading
decay $K^+ \rightarrow \pi^0 e^+ \nu$, the resulting theoretical
expressions for Br( $K_L\to\pi^0\nu\bar\nu$) and Br($K^+\to\pi^+\nu\bar\nu$)
  are only functions of the CKM parameters, the QCD scale
 $\Lms$
 and the
quark masses $\mt$ and $\mc$. The isospin braking corrections calculated in
\cite{MP} reduce the $K^+$ and $K_L$ branching ratios by $10\%$ and $5.6\%$
respectively.
The long distance contributions to
$K^+ \rightarrow \pi^+ \nu \bar{\nu}$ have been
considered in \cite{RS} and found to be very small: a few percent of the
charm contribution to the amplitude at most, which is safely neglegible.
The long distance contributions to $K_L\to\pi^0\nu\bar\nu$ are negligible
as well. 

The explicit expressions for $Br(K^+ \rightarrow \pi^+ \nu \bar{\nu})$ 
and $Br(K_L\to\pi^0\nu\bar\nu)$ can be found in \cite{BBL}. Here we
give approximate expressions in order to exhibit various dependences:

\begin{equation}\label{bkpn}
Br(K^+ \rightarrow \pi^+ \nu \bar{\nu})=
0.7\cdot 10^{-10}\left[\left [ \frac{|V_{td}|}{0.010}\right ]^2
\left [\frac{\mid V_{cb}\mid}{0.040} \right ]^2
\left [\frac{\mtb(\mt)}{170~\gev} \right ]^{2.3} 
 +{\rm cc+tc}\right]
\end{equation}

\begin{equation}\label{bklpn}
Br(K_L\to\pi^0\nu\bar\nu)=
2.8\cdot 10^{-11}\left [ \frac{\eta}{0.37}\right ]^2
\left [\frac{\mtb(\mt)}{170~GeV} \right ]^{2.3} 
\left [\frac{\mid V_{cb}\mid}{0.040} \right ]^4 
\end{equation}
where in (\ref{bkpn}) we have shown explicitly only the pure top
contribution.

The impact of NLO calculations is best illustrated by giving the
scale uncertainties in the leading order and after the inclusion
of the next-to-leading corrections:
\begin{equation}
Br(K^+ \rightarrow \pi^+ \nu \bar{\nu})=
(1.00\pm0.22)\cdot 10^{-10}\quad =>\quad
(1.00\pm0.07)\cdot 10^{-10}
\end{equation}
\begin{equation}
Br(K_L\to\pi^0\nu\bar\nu)=(3.00\pm0.30)\cdot 10^{-11}\quad =>\quad
(3.00\pm0.04)\cdot 10^{-11}
\end{equation}
The reduction of the scale uncertainties is truly impressive.
The reduction of the scale uncertainty in 
$Br(K^+ \rightarrow \pi^+ \nu \bar{\nu})$ corresponds to the reduction
in the uncertainty in the determination of $|V_{td}|$ from $\pm 14\%$ to 
$\pm 4\%$.

Scanning the input parameters of table 1 one finds \cite{BJL96b}:
\begin{equation}
Br(K^+ \rightarrow \pi^+ \nu \bar{\nu})=
(9.1\pm 3.2)\cdot 10^{-11}\quad,\quad
Br(K_L\to\pi^0\nu\bar\nu)=(2.8\pm 1.7)\cdot 10^{-11}
\end{equation}
where the errors come dominantly from the uncertainties in the
CKM parameters. 

The present experimental bound on $Br(K^+\to \pi^+\nu\bar\nu)$
is $2.4 \cdot 10^{-9}$ \cite{Adler95}. A new bound $ 2 \cdot 10^{-10}$ for 
this decay  is expected from E787 at AGS in Brookhaven in 1997.
In view of the clean character of this decay a measurement of its
branching ratio at this level would signal the presence of physics
beyond the standard model. Further experimental improvements for
this branching ratio are discussed by Littenberg in these proceedings 
and in \cite{Cooper}.
The present upper bound on $Br(K_L\to \pi^0\nu\bar\nu)$ from
FNAL experiment E731 \cite{WEAVER} is $5.8 \cdot 10^{-5}$. 
FNAL-E799 expects to reach
the accuracy ${\cal O}(10^{-8})$ and a very interesting new proposal 
 AGS2000 \cite{AGS2000} 
expects to reach the single event sensitivity $2\cdot 10^{-12}$
allowing a $10\%$ measurement of the expected branching ratio. 
It is hoped that also JNAF(CEBAF), KAMI and KEK will make efforts
to measure this gold-plated  decay. Such measurements will also put
constraints on the physics beyond the standard model \cite{Sher}.
We will return to both decays in section 6.

\subsection{$K_L\to \mu\bar\mu$}
The rare decay $K\to\mu\bar\mu$ is CP conserving and in addition to its
short-distance part, given by Z-penguins and box diagrams, receives 
important contributions from the
two-photon intermediate state, which are difficult to calculate
reliably \cite{gengng:90,belangergeng:91,ko:92,Singer,eeg} as discussed
by Eeg at this workshop.

This latter fact is rather unfortunate because the
short-distance part is, similarly to $K\to\pi\nu\bar\nu$, free of hadronic
uncertainties and if extracted from the existing data would give a useful
determination of the Wolfenstein parameter $\varrho$.
 The separation
of the short-distance piece from the long-distance piece in the measured
rate is very difficult however.

The analysis of the short distance part proceeds in essentially the same
manner as for $K\to \pi\nu\bar\nu$. The only difference enters through 
the lepton line in the box contribution which makes the $m_t$ dependence
stronger and $m_c$ contribution smaller. 
The next-to-leading QCD corrections to this decay
have been calculated in \cite{BB13}. This calculation
reduced the theoretical uncertainty
due to the choice of the renormalization scales present in the
leading order expressions from $\pm 24\%$ to $\pm 10\%$. The
remaining scale uncertainty which is larger than in 
$K^+ \rightarrow \pi^+ \nu \bar{\nu}$ is related to a particular
feature of the perturbative expansion in this decay \cite{BB13}.
An approximate expression for the short distance part is given as
follows:
\begin{equation}\label{bkmumu}
Br(K_L\to\mu\bar\mu)_{SD}=
0.9\cdot 10^{-9}\left (1.2 - \bar\varrho \right )^2
\left [\frac{\mtb(\mt)}{170~GeV} \right ]^{3.1} 
\left [\frac{\mid V_{cb}\mid}{0.040} \right ]^4 
\end{equation}
In the absence of charm contribution, "1.2" in the first parenthesis
would be replaced by "1.0".

Scanning the input parameters of table 1 we find :
\begin{equation}\label{BSD}
Br(K_L\to\mu\bar\mu)_{SD}=
(1.3\pm 0.6)\cdot 10^{-9}
\end{equation}
where the error comes dominantly from the uncertainties in the
CKM parameters. 

Now the full branching ratio can be written generally as follows:
\begin{equation}\label{BSDa}
Br(K_L\to\mu\bar\mu)= |\RE A|^2+|\IM A|^2\qquad
\RE A = A_{SD}+A_{LD}
\end{equation}
with $\RE A$ and $\IM A$ denoting the dispersive and absorptive contributions
respectively. The absorptive contribution can be calculated using the
data for $K_L\to \gamma\gamma$ and is known under the name of the unitarity
bound \cite{Unitary}. 
One finds $(6.81\pm 0.32)\cdot 10^{-9}$ which is very close to the
experimental measurements

\begin{equation}
Br(K_L\to \bar\mu\mu) =\left\{ \begin{array}{ll}
(6.86\pm0.37)\cdot 10^{-9}~({\rm BNL} 791) & \cite{PRINZ} \\
(7.9\pm 0.6 \pm 0.3)\cdot 10^{-9}~({\rm KEK} 137) & 
\cite{Akagi} \end{array} \right.
\end{equation}
which give the world average:
\begin{equation}\label{princ}
Br(K_L\to \bar\mu\mu) = (7.1\pm 0.3)\cdot 10^{-9}
\end{equation}
The accuracy of this result is impressive $(\pm 4\%)$. It will be
reduced to $(\pm 1\%)$ at BNL in the next years.

The BNL791 group using their data and the unitarity bound extracts
$|\RE A|^2\le 0.6\cdot 10^{-9}$ at $90\%$ C.L. This is a bit lower than the
short distance prediction in (\ref{BSD}). Unfortunately in order to use
this result for the determination of $\varrho$ the long distance dispersive 
part $A_{LD}$ resulting from the intermediate off-shell two photon states 
should be known.
The present estimates of $A_{LD}$ are too uncertain to obtain a useful
information on $\varrho$. It is believed that the measurement of
$Br(K_L\to e\bar e \mu\bar\mu)$ should help in estimating this part.
The present result $(2.9+6.7-2.4)\cdot 10^{-9}$ from E799 
should therefore be improved.

More details on this decay can be found in \cite{PRINZ,BB13,CPRARE,Singer}
and in the talk by Eeg at this workshop.
More promising from theoretical point of view is the parity-violating
asymmetry in $K^+\to \pi^+\mu^+\mu^-$ \cite{GENG,BB5}.          
Finally as stressed by Pich at this workshop, the longitudinal
polarization in this decay is rather sensitive to contributions
beyond the standard model \cite{EKPICH}.

\subsection{Classification}
It is probably a good idea to end this section by grouping various decays and 
quantities into four distinct classes with respect to theoretical
uncertainties. I include in this
classification also B-decays and in particular CP asymmetries in B decays
which I will briefly discuss in the following section.
\subsubsection{Gold-Plated Class}
These are the decays with essentially no theoretical uncertainties:
\begin{itemize}
\item
CP asymmetries in $B_d\to \psi K_S$ and  $B_s\to\psi\phi$ which
measure the angle $\beta$ and the parameter $\eta$ respectively,
\item
The ratio  $Br(B\to X_d\nu\bar\nu)/ Br(B\to X_s\nu\bar\nu)$
which offers the cleanest direct determination of the ratio 
$|V_{td}/V_{ts}|$,
\item
Rare K-decays $K_L \to \pi^0\nu\bar\nu$ and $K^+\to \pi^+\nu\bar\nu$
which offer very clean determinations of $\IM\lambda_t (\eta)$ and
$|V_{td}|$ respectively.
\end{itemize}
\subsubsection{Class 1}
\begin{itemize}
\item
CP asymmetry in $B^0\to \pi^+\pi^-$ relevant for the angle $\alpha$ and the
CP asymmetries in $B^{\pm}\to D_{CP}K^{\pm}$,
$B_s\to D_sK$ and $B^0\to \bar D^0K^*$ all relevant for the angle $\gamma$.
These CP asymmetries
require additional strategies in order to determine these angles without
hadronic uncertainties.
\item
Ratios $Br(B_d\to l\bar l)/Br(B_s\to l\bar l)$ and
$(\Delta M)_d/(\Delta M)_s$
which give  good measurements of $|V_{td}/V_{ts}|$ 
provided the SU(3) breaking effects in the ratios $F_{B_d}/F_{B_s}$
and $\sqrt{B_d}F_{B_d}/\sqrt{B_s}F_{B_s}$ can be brought under control.
\end{itemize}
\subsubsection{Class 2}
Here I group  quantities or decays with presently moderete or substantial
theoretical uncertainties which should be considerably reduced in the next
five years. In particular I assume that the uncertainties in $B_K$
and $\sqrt{B}F_B$ will be reduced below 10\%.
\begin{itemize}
\item
$B\to X_{s,d}\gamma$, $B\to X_{s,d} e^+ e^-$, $B\to K^*(\rho)e^+e^-$
\item
$(\Delta M)_d$, $(\Delta M)_s$, $\vcb_{excl}$, 
$\vcb_{incl}$,
$| V_{ub}/V_{cb}|_{incl}$
\item
Some CP asymmetries in B-decays reviewed for instance in \cite{B95}
\item
$\varepsilon_K$ and $K_L\to \pi^0 e^+e^-$
\end{itemize}
\subsubsection{Class 3}
Here we have a list of important decays with large theoretical
uncertainties which can only be removed by a dramatic progress
in non-perturbative techniques:
\begin{itemize}
\item
CP asymmetries in most $B^{\pm}$-decays
\item
$B_d\to K^*\gamma$, Non-leptonic B-decays, $| V_{ub}/V_{cb}|_{excl}$
\item
$\varepsilon^{\prime} /\varepsilon$, $K\to \pi\pi$, $\Delta M(K_L-K_s)$,
$K_L\to\mu\bar\mu$, hyperon decays and so on.
\end{itemize}
It should be stressed that even in the presence of theoretical
uncertainties a measurement of a non-vanishing 
ratio $\varepsilon^{\prime}/\varepsilon$ or a non-vanishing CP asymmetry
in charged B-decays would signal direct CP violation excluding
superweak scenarios \cite{WO1}. This is not guaranteed by
several clean decays of the gold-plated class or class 1 \cite{WIN} except for 
$B^{\pm}\to D_{CP} K^{\pm}$.

\section{CP-B Asymmetries versus $K \to \pi \nu\bar\nu$}
\subsection{CP-Asymmetries in B-Decays}
CP violation in B decays is certainly one of the most important 
targets of B factories and of dedicated B experiments at hadron 
facilities. It is well known that CP violating effects are expected
to occur in a large number of channels at a level attainable at 
forthcoming experiments. Moreover there exist channels which
offer the determination of CKM phases essentially without any hadronic
uncertainties. Since CP violation in B decays has been already
reviewed in two special talks by Nir and Nakada at this workshop 
and since in addition extensive reviews can be found in the literature
\cite{NQ,B95,RFD}, let me concentrate only on the most important points.

The CP-asymmetry in the decay $B_d^0 \rightarrow \psi K_S$ allows
 in the standard model
a direct measurement of the angle $\beta$ in the unitarity triangle
without any theoretical uncertainties \cite {BSANDA}.
 Similarly the CP asymmetry in the decay
$B_d^0 \rightarrow \pi^+ \pi^-$ gives the angle $\alpha$, although
 in this case strategies involving
other channels are necessary in order to remove hadronic
uncertainties related to penguin contributions
\cite{CPASYM}.

We have then for the time-dependent asymmetries
\begin{equation}\label{113c}
 A_{CP}(\psi K_S,t)=-\sin((\Delta M)_d t)\sin(2\beta) 
\qquad
   A_{CP}(\pi^+\pi^-,t)=-\sin((\Delta M)_d t)\sin(2\alpha+\theta_P) 
\end{equation}
where $\theta_P$ represents the "QCD penguin pollution" which has to be
taken care of in order to extract $\alpha$. The most popular strategy
is the isospin analysis of Gronau and London \cite{CPASYM}. It
requires however the measurement of $Br(B^0\to \pi^0\pi^0)$ which is
expected to be below $10^{-6}$: a very difficult experimental task.
For this reason other strategies avoiding this channel or estimating
the size of the penguin contribution have been proposed. Since this
is the K-physics workshop I will not review them here and refer to
 recent reviews \cite{B95,Gronau,RFD} where various alternative strategies 
for
the determination of $\alpha$ and the issues of the determination of
the angle $\gamma$ in the decays of class 1 \cite{Wyler,DUN2,adk}
are discussed.

In what follows let us assume that the problems with the determination
of $\alpha$ will be solved somehow. Since in the usual unitarity triangle
 one side is known, it suffices to measure
two angles to determine the triangle completely. This means that
the measurements of $\sin 2\alpha$ and $\sin 2\beta$ can determine
the parameters $\varrho$ and $\eta$.
As the standard analysis of the unitarity triangle of section 3
shows, $\sin(2\beta)$ is expected to be large: $\sin(2\beta)=0.58\pm 0.22$
implying the integrated asymmetry  $A_{CP}(\psi K_S)$
as high as $(30 \pm 10)\%$.
The prediction for $\sin(2\alpha)$ is very
uncertain on the other hand $(0.1\pm0.9)$ and even a rough measurement
of $\alpha$ would have a considerable impact on our knowledge of
the unitarity triangle as stressed in \cite{BLO} and recently in
\cite{BB96}.
\subsection{UT from CP-B and $K\to\pi\nu\bar\nu$ }
Let us then compare the potentials of the CP asymmetries in
determining the parameters of the standard model with those
of the cleanest rare K-decays: $K_L\to\pi^0\nu\bar\nu$ and
$K^+\to\pi^+\nu\bar\nu$ \cite{BB96}.
Measuring $\sin 2\alpha$ and $\sin 2\beta$ from CP asymmetries in
$B$ decays allows, in principle, to fix the 
parameters $\bar\eta$ and $\bar\varrho$, which can be expressed as
\cite{B94}
\begin{equation}\label{ersab}
\bar\eta=\frac{r_-(\sin 2\alpha)+r_+(\sin 2\beta)}{1+
  r^2_+(\sin 2\beta)}\qquad
\bar\varrho=1-\bar\eta r_+(\sin 2\beta)
\end{equation}
where $r_\pm(z)=(1\pm\sqrt{1-z^2})/z$.
In general the calculation of $\bar\varrho$ and $\bar\eta$ from
$\sin 2\alpha$ and $\sin 2\beta$ involves discrete ambiguities.
As described in \cite{B94}
they can be resolved by using further information, e.g. bounds on
$|V_{ub}/V_{cb}|$, so that eventually the solution (\ref{ersab})
is singled out.
\\
Alternatively, $\bar\varrho$ and $\bar\eta$ may also be determined
from $K^+\to\pi^+\nu\bar\nu$ and $K_L\to\pi^0\nu\bar\nu$ alone
\cite{BH,BB4}. An interesting feature of this possibility is in
particular that the extraction of $\sin 2\beta$ from these
two modes is essentially independent of $\mt$ and $V_{cb}$
\cite{BB4}. This fact enables a rather accurate determination of
$\sin 2\beta$ from $K\to\pi\nu\bar\nu$.

A comparison of both strategies
is displayed in Table \ref{tabkb}, where 
the following input has been used
\begin{equation}\label{vcbmt}
|V_{cb}|=0.040\pm 0.002\qquad m_t=(170\pm 3) GeV
\end{equation}
\begin{equation}\label{bklkp}
B(K_L\to\pi^0\nu\bar\nu)=(3.0\pm 0.3)\cdot 10^{-11}\qquad
B(K^+\to\pi^+\nu\bar\nu)=(1.0\pm 0.1)\cdot 10^{-10}
\end{equation}
\\
The measurements of CP asymmetries in $B_d\to\pi\pi$ and
$B_d\to J/\psi K_S$, expressed in terms of $\sin 2\alpha$ and
$\sin 2\beta$, are taken to be
\begin{equation}\label{sin2a2bI}
\sin 2\alpha=0.40\pm 0.10 \qquad \sin 2\beta=0.70\pm 0.06
\qquad ({\rm scenario\ I})
\end{equation}
\begin{equation}\label{sin2a2bII}
\sin 2\alpha=0.40\pm 0.04 \qquad \sin 2\beta=0.70\pm 0.02
\qquad ({\rm scenario\ II})
\end{equation}
Scenario I corresponds to the accuracy being aimed for at $B$-factories
and HERA-B prior to the LHC era. An improved precision can be anticipated from
LHC experiments, which we illustrate with our choice of scenario II.

As can be seen in Table \ref{tabkb}, the CKM determination
using $K\to\pi\nu\bar\nu$ is competitive with the one based
on CP violation in $B$ decays, except for $\bar\varrho$ which
is less constrained by the rare kaon processes.
\begin{table}
\begin{center}
\begin{tabular}{|c||c|c|c|}\hline
&$K\to\pi\nu\bar\nu$&$B\to\pi\pi, J/\psi K_S$ (I) 
&$B\to\pi\pi, J/\psi K_S$ (II) \\
\hline
\hline
$|V_{td}|/10^{-3}$&$10.3\pm 1.1(\pm 0.9)$&$8.8\pm 0.5(\pm 0.3)$ 
&$8.8\pm 0.5(\pm 0.2)$ \\
\hline
$|V_{ub}/V_{cb}|$&$0.089\pm 0.017(\pm 0.011)$
&$0.087\pm 0.009(\pm 0.009)$&$0.087\pm 0.003(\pm 0.003)$ \\
\hline 
$\bar\varrho$&$-0.10\pm 0.16(\pm 0.12)$&$0.07\pm 0.03(\pm 0.03)$
&$0.07\pm 0.01(\pm 0.01)$ \\
\hline
$\bar\eta$&$0.38\pm 0.04(\pm 0.03)$&$0.38\pm 0.04(\pm 0.04)$
&$0.38\pm 0.01(\pm 0.01)$ \\
\hline
$\sin 2\beta$&$0.62\pm 0.05(\pm 0.05)$&$0.70\pm 0.06(\pm 0.06)$
&$0.70\pm 0.02(\pm 0.02)$ \\
\hline
${\rm Im}\lambda_t/10^{-4}$&$1.37\pm 0.07(\pm 0.07)$
&$1.37\pm 0.19(\pm 0.15)$&$1.37\pm 0.14(\pm 0.08)$ \\
\hline
\end{tabular}
\end{center}
\caption[]{Illustrative example of the determination of CKM
parameters from $K\to\pi\nu\bar\nu$ and from CP violating
asymmetries in $B$ decays. The relevant input is as described
in the text. Shown in brackets are the errors one obtains
using $V_{cb}=0.040\pm 0.001$ instead of $V_{cb}=0.040\pm 0.002$.
\label{tabkb}}
\end{table}
On the other hand ${\rm Im}\lambda_t$ is better determined
in the kaon scenario. It can be obtained from
$K_L\to\pi^0\nu\bar\nu$ alone and does not require knowledge
of $V_{cb}$ which enters ${\rm Im}\lambda_t$ when derived
from $\sin 2\alpha$ and $\sin 2\beta$.
This analysis suggests that $K_L\to\pi^0\nu\bar\nu$ should eventually 
yield the most accurate value of ${\rm Im}\lambda_t$.
This would be an important result since ${\rm Im}\lambda_t$
plays a central role in the phenomenology of CP violation
in $K$ decays and is furthermore equivalent to the 
Jarlskog parameter $J_{CP}$ \cite{CJ}, 
the invariant measure of CP violation in the Standard Model, 
$J_{CP}=\lambda(1-\lambda^2/2){\rm Im}\lambda_t$.

There is another virtue of the comparision of the determinations
of various parameters using CP-B asymmetries with the determinations
in very clean decays $K\to\pi\nu\bar\nu$. Any substantial deviations
from these two determinations would signal new physics beyond the
standard model.

On the other hand  
unprecedented precision for all basic CKM
parameters could be achieved by combining the cleanest K and 
B decays \cite{B94} . 
While $\lambda$ is obtained as usual from
$K\to\pi e\nu$, $\bar\varrho$ and $\bar\eta$ could be determined
from $\sin 2\alpha$ and $\sin 2\beta$ as measured in CP
violating asymmetries in $B$ decays. Given $\eta$, one could
take advantage of the very clean nature of $K_L\to\pi^0\nu\bar\nu$
to extract $A$ or, equivalently $|V_{cb}|$. This determination
benefits further from the very weak dependence of $|V_{cb}|$ on
the $K_L\to\pi^0\nu\bar\nu$ branching ratio, which is only with
a power of $0.25$. Moderate accuracy in $B(K_L\to\pi^0\nu\bar\nu)$
would thus still give a high precision in $|V_{cb}|$.
As an example we take $\sin 2\alpha=0.40\pm 0.04$,
$\sin 2\beta=0.70\pm 0.02$ and 
$B(K_L\to\pi^0\nu\bar\nu)=(3.0\pm 0.3)\cdot 10^{-11}$,
$m_t=(170\pm 3)GeV$. 
This yields
\begin{equation}\label{rhetvcb}
\bar\varrho=0.07\pm 0.01\qquad
\bar\eta=0.38\pm 0.01\qquad
|V_{cb}|=0.0400\pm 0.0013
\end{equation}
which would be a truly remarkable result. Again the comparision of
this determination of $|V_{cb}|$ with the usual one in tree level
B-decays would offer an excellent test of the standard model
and in the case of discrepancy would signal physics beyond the
standard model.  

\section{A Look beyond the Standard Model}
In this  review we have concentrated on rare decays and  CP violation in the
standard model. The structure of rare decays and of CP violation in 
extensions of the 
standard model may deviate from this picture.
Consequently the situation in this field could turn out to be very different
from the one presented here. It is appropriate then to end this review with
a few remarks on the physics beyond the standard model. Much more
elaborate discussion can be found in the talk of Nir presented
at this workshop and in \cite{NIRNEW,Ligetti}.
\subsection{Impact of New Physics}
There is essentially no impact on $|V_{us}|$, $|V_{cb}|$ and $|V_{ub}|$
determined in tree level decays. This is certainly the case for the first
two elements. In view of the smallness of $|V_{ub}|$ a small impact from
the loop contributions (sensitive to new physics) to leading decays
could in principle be present. However in view of many theoretical
uncertainties in the determination of this element such contributions
can be safely neglected at present.

There is in principle a substantial impact of new physics on the
determination of $\varrho$, $\eta$, $|V_{td}|$, $\IM\lambda_t$ and
generally on the unitarity triangle through the loop induced
decays which can receive new contributions from internal chargino, 
charged Higgs, stops, gluinos and other exotic exchanges. If the
quark mixing matrix has the CKM structure, the element $|V_{ts}|$
on the other hand will be only slightly affected by these new
contributions. Indeed from the unitarity of the CKM matrix
$|V_{ts}|/|V_{cb}|=1-{\cal O}(\lambda^2)$ and the new contributions
could only affect the size of the ${\cal O}(\lambda^2)$ terms which
amounts to a few percent at most. This situation makes the study
of new physics in rare B decays governed by $|V_{ts}|$ somewhat
easier than in rare K-decays and B decays which are governed by
$|V_{td}|$. Indeed in the latter decays the impact of new physics
is felt both in the CKM couplings and in the $\mt$ dependent functions,
which one has to disantangle, whereas in the former decays
mainly the impact of new physics on the $m_t$ dependent functions
is felt.

Similarly if no new phases in the quark mixing are present, 
the formulae for CP asymmetries in B-decays remain unchanged and these
asymmetries measure again the phases of the CKM matrix as in the
standard model. Thus even if there is some new physics in the loop
diagrams we will not see it in the clean asymmetries directly if
there are no new phases in the quark mixing matrix. In order
to search for new physics  the comparision of the values of CKM phases
determined from CP asymmetries and from loop induced decays is then mandatory.

The situation becomes more involved if the quark mixing involves
more angles and new phases and in addition there are new parameters 
in the Higgs, SUSY and generally new physics sector. For instance
in such a case the "gold-plated" asymmetry in $B \to\psi K_S$ 
would take the form \cite{NIRNEW}:
\begin{equation}\label{113u}
 A_{CP}(\psi K_S,t)=-\sin((\Delta M)_d t)\sin(2\beta+\theta_{NEW}) 
\end{equation}
implying that not $2\beta$ but  $2\beta+\theta_{NEW}$ is measured by
the asymmetry.
 
This short discussion makes it clear that in order to search effectively
for new physics it is essential to measure and calculate as many
processes and compare the resulting CKM parameters with each other.
Graphically this corresponds simply to figure 2. In this enterprise
the crucial role will be played by very clean decays of the "gold-plated"
class and of classes 1 and possibly 2 in which the new physics will
not be hidden by theoretical uncertainties present in the decays of class 3.

\subsection{Signals of New Physics}
New Physics will be signaled in principle in various ways. Here are some
obvious examples:
\begin{itemize}
\item
Standard model predictions for various branching ratios and CP
asymmetries will disagree with data,
\item
$(\varrho,\eta)$  determined in K-physics will disagree with
$(\varrho,\eta)$ determined in B-physics,
\item
$(\varrho,\eta)$ determined in loop induced decays will disagree
with $(\varrho,\eta)$ determined through CP asymmetries,
\item
Forbidden and very rare decays will occur at unexpected level:
$K_L\to \mu e$, $K\to \pi \mu e$, $d_N$, $d_e$, $D^0-\bar D^0$
mixing, CP violation in D-decays \cite{Burdman} etc.,
\item
Unitarity Triangle will not close.
\end{itemize}
\subsection{General Messages}
Let us end this discussion with some general messages on New Physics which
have been stressed in particular by Nir at this workshop.

As discussed by Gavela at this workshop, baryogenesis suggests that there 
is CP violation
outside the Standard Model. The single CKM phase simply does not give
enough CP Violation for the required baryon asymmetry \cite{Gavela}. 
It is however not unlikely that large new sources of
CP violation necessary for baryogenesis could be present at the electroweak 
scale \cite{Nelson}. They 
are present for instance in general SUSY models and in multi-Higgs models.

It should be stressed that baryogenesis and the required additional CP
violation being flavour diagonal may have direct impact on the electric
dipole moments but have no direct impact on FCNC processes. However
new physics required for baryon asymmetry could bring new phases
relevant for FCNC.

Concentrating on SUSY for a moment, more general and natural 
SUSY models give
typically very large CP violating effects \cite{Gerar} and FCNC transitions
 \cite{Nilles} which
are inconsistent with the experimental values of $\varepsilon_K$, 
$K_L-K_S$ mass difference and the bound on the electric dipole moment
of the neutron. In order to avoid such problems, special forms of 
squark mass matrices \cite{Masiero}
and fine tunning of phases are necessary. In addition
one frequently assumes that CP violation and FCNC are absent at tree
level. 
In the limiting case one ends with a special version of 
the MSSM in which to a good
approximation CP violation and FCNC processes are governed by the
CKM matrix and the new effects are dominantly described by loop diagrams
with internal stop, charginos and charged higgs exchanges \cite{FRERE}.
It is then not surprising that in the quark sector new effects in MSSM 
compared with SM predictions for FCNC transitions 
are rather moderate, although for a particular
choice of parameters and certain quantities still enhancements 
(or suppressions)
by factors 2-3 cannot be excluded \cite{Branco}. 
Larger effects are expected in the lepton
sector and in electric dipole moments. Similar comments about the size
of new effects apply to multi-higgs models and left-right symmetric
models.

Large effects are still possible in models with tree level FCNC
transitions, leptoquarks, models with horizontal gauge symmetries,
technicolour and top-colour models \cite{NIRNEW,Ligetti}. 
Unfortunately these 
models
contain many free parameters and at present the only thing  one
can do is to bound numerous new couplings and draw numerous
curves which from my point of view is not very exciting.

On the other hand it is to be expected that  clearest signals
of new physics may come precisely from very exotic physics
which would cause the decays $K_L\to \mu e$, $K\to \pi \mu e$, 
T-violating $\mu$-polarization in $K^+\to \pi^0\mu^+\nu$ to occur. Also
sizable values of $d_N$, $d_e$, $D^0-\bar D^0$
mixing and of CP violation in D-decays and top decays are very interesting
in this respect.

It should however be  stressed once more that theoretically
cleanest decays belonging to the top classes of section 5 will
certainly play important roles in the search for new physics
and possibly will offer its first signals.
\section{Outlook}
{\bf There is clearly an excting time ahead of us !}
\section{Final Remarks}
This was a very enjoyable workshop brilliantly organized by the team
lead by Lydia and Louis Fayard. In particular the interactions between
theorists and experimentalists were very fruitful. Even if the first
and the last word have been given to experimentalists 
(Rene Turlay, Bruce Winstein), the first and the last chairmanship
have been given to theorists (Eduardo de Rafael, Fred Gilman).
I do hope very much that these fruitful interactions between experimentalists
and theorists will continue so that in the year 1999, when hopefully
Lydia and Louis will organize another K-physics workshop in Orsay, we will
all agree that $\epe\not=0$ and that the main K-physics target for
the next decade should be the measurement of $K_L\to \pi^0\nu\bar\nu$.

{\bf Acknowledgements}: I would like to thank Markus Lautenbacher for
help in producing the figures and him, Gerhard Buchalla and Matthias
Jamin for wonderful collaboration. This work has been supported by the
German Bundesministerium f\"ur Bildung and Forschung under contract 
06 TM 743 and DFG Project Li 519/2-1.

\vfill\eject

\end{document}